\pdfoutput=1
\documentclass[12pt]{article}

\usepackage{times}
\usepackage{multirow}
\usepackage{amssymb}
\usepackage{amsmath}
\usepackage{graphicx}
\usepackage[]{setspace}
\usepackage{tocloft}
\usepackage{cite}

\newenvironment{sciabstract}{%
\begin{quote} \bf}
{\end{quote}}

\title{Revealing the Anatomy of Vote Trading\footnote{We acknowledge the financial support of the Institute for New Economic Thinking at the Oxford Martin School through the grant INET12-9001, and a seed grant provided by the University of Basel Research Found as well as support from the Swiss National Science Foundation (grant 168848). We are grateful to Patrick Balles, Irene D\'avalos, Andrew Elliot, Teppo Felin, Steve Fisher, Bernard Grofman, Robert Hahn, Jeffrey Lienert, Eduardo Lopez, Dietmar Maringer, Armando Meier, Reto Odermatt, Sanna Ojanper\"a, Dennis Quinn, Felix Reed-Tsochas, Matteo Richiardi, Serguei Saavedra, Kurt Schmidheiny, Nicolas Schreiner, Michaela Slotwinski, Alois Stutzer, and conference participants at the International Conference for Computational Social Science 2015, the 2016 Meeting of the European Public Choice Society, the SSES Congress 2016, as well as seminar participants at Harvard University, the University of Oxford, the University of Zurich, the University of Basel and the University of Fribourg  for helpful remarks.}}

\author{Omar A. Guerrero,$^{1}$ Ulrich Matter,$^{2}$\\
\\
\small{$^{1}$CABDyN Complexity Centre, Sa\"id Business School, University of Oxford;}
\\
\small{Institute for New Economic Thinking at the Oxford Martin School;}
\\
\small{Department of Computer Science, Aalto University.}
\\
\small{$^{2}$Berkman Klein Center for Internet \& Society, Harvard University;}\\
\small{Faculty of Business and Economics, University of Basel.}
\\
\small{$^\ast$E-mail: omar.guerrero@sbs.ox.ax.uk}
}

\date{}

\begin{document} 

\baselineskip24pt

\maketitle 

{\centering
  Supplementary material is available upon request.\par
}

\begin{sciabstract}
\begin{singlespace}
Cooperation in the form of vote trading, also known as logrolling, is central for law-making processes, shaping the development of democratic societies. Empirical evidence of logrolling is scarce and limited to highly specific situations because existing methods are not easily applicable to broader contexts. We have developed a general and scalable methodology for revealing a network of vote traders, allowing us to measure logrolling on a large scale. Analysis on more than 9 million votes spanning 40 years in the U.S. Congress reveals a higher logrolling prevalence in the Senate and an overall decreasing trend over recent congresses, coincidental with high levels of political polarization. Our method is applicable in multiple contexts, shedding light on many aspects of logrolling and opening new doors in the study of hidden cooperation.
\end{singlespace}
\end{sciabstract}

\clearpage

\section*{Introduction}

The law-making process of any democratic society is critical to its socioeconomic development. For example, in 2013, a political gridlock in the U.S. Congress led to a government shutdown with an estimated economic cost of at least 0.3\% of the quarterly GDP \cite{_gross_2014}. In this case, cooperation between legislators with opposing political views was necessary to end this blockage. However, in other situations, this kind of cooperation might just as well favor special interest groups and thereby hurt third parties. Here, we look at vote trading, or logrolling, as one such type of cooperative behavior in which two legislators vote in favor of each others' preferred bills in order to secure their passage. While it is extensively discussed in theoretical literature from Economics, Game Theory, and Political Science \cite{buchanan_calculus_1962, wilson_axiomatic_1969,tullock_simple_1970,riker_paradox_1973,bernholz_stability_1978,casella_vote_2014}, empirical evidence on vote trading is scarce and limited to the study of a few specific votes or legislators. Up to now we have lacked the tools with which to study the prevalence of logrolling on a large scale. Insights into the relative importance of this hidden cooperation in legislative assemblies, however, are crucial in order to understand the success and failure of alternative institutional arrangements used to address economic and social problems. Leaving aside value judgements about the benefits, costs, and ethics of this type of cooperation, all social sciences dedicated to the study of politics consider vote trading -- on both theoretical grounds and anecdotal evidence -- a central aspect of democratic processes (not only in legislative assemblies, but also in any group making decisions with the simple majority rule). However, without a widely applicable general approach that allows vote trading to be captured empirically, our understanding of its determinants, mechanisms, and consequences is limited.

The challenge of empirically assessing the prevalence of logrolling lies in its nature. Legislators on both sides of a vote-trading agreement are keen to keep such a deal secret. Strategically offering one's vote to a fellow legislator with opposing political views can contrast starkly with the expectation of voters and peers with similar political views. Hence, empirically capturing vote trading requires us to measure something that is not directly observable. Previous empirical studies that have tried to detect or measure vote trading focused on a few specific bills \cite{stratmann_effects_1992,stratmann_logrolling_1995} or covered trades within a small group of legislators \cite{cohen_friends_2014}. These approaches are rather designed for and very useful in analyzing highly specific aspects of vote trading. Because they require extensive knowledge about the sociopolitical context of specific bills and legislators, they are unfit for measuring logrolling on a large scale. We have developed a novel approach to measuring the prevalence of vote trading by combining new insights from complex networks literature with computational methods and well-established theory on logrolling.

When seen as a systemic phenomenon, vote trading is driven by fundamental incentives set by a common democratic procedure: voting under the simple majority rule. As legislators have only one vote per decision, this rule cannot take into account preference intensities for different bills (that is, legislators cannot weight their vote on a bill depending on whether they particularly favor or dislike it). They simply have one vote per issue. However, legislators can give up their preferred position on a bill and `offer' their vote in order to get support in the passage of their favorite bills. Vote trading is thus mutually beneficial for legislators who are particularly interested in the passage of specific bills. It also implies that when trading their votes, legislators do not vote sincerely, i.e., in line with their preferred policy position. Importantly, offering to vote in favor of bills one dislikes in order to get support for one's most favored bill(s) primarily makes sense in the context of narrow voting outcomes \cite{stratmann_effects_1992,cohen_friends_2014}. In narrow outcomes, the marginal support is particularly valuable for the supporters of the bill. With higher incentives to trade votes, roll calls in narrow outcomes do not necessarily reveal legislators' actual political positions on an issue but could be the result of exchanging favors. Conversely, legislators' observed decisions in roll calls decided by a wide margin are more likely to reveal their preferred policy positions (with a few exceptions such as protest voting \cite{asch_studies_1956}). In roll calls, legislators engaged in vote trading can observe each others' actions once the votes are cast and thus know with certainty whether their partner kept his or her part of the bargain. However, as they have strong incentives to keep such deals secret, they cannot directly punish defection (e.g., publicly blame or sue a legislator who does not keep his or her part of the bargain). Cooperation in the form of logrolling is thus only likely to evolve as reciprocal behavior over repeated interactions \cite{axelrod_evolution_1984}. With this framework in mind, we have developed a methodology that measures logrolling on large-scale roll call datasets. The theoretical pillars of the method can be summarized in four principles that have been tested in specific voting contexts \cite{stratmann_effects_1992,stratmann_logrolling_1995}:

\begin{enumerate}
    \item Incentives to trade votes are stronger the \emph{narrower} vote outcomes are.
    \item If a legislator trades a vote, then he or she votes in the opposite direction to what would be predicted, i.e. the legislator \emph{deviates} from his or her preferred position. 
    \item A deviation is considered a potential trade if it is \emph{directed} to benefit a legislator with a clear interest in passing the bill that is being voted on.
    \item Directed deviations are considered traded votes only if they are \emph{reciprocal} and \emph{mutually profitable}.
\end{enumerate}

In a nutshell, our method consists of four components: estimating legislators' preferred policy positions; detecting deviations from those positions; identifying the beneficiaries of those deviations; and measuring reciprocity between deviators and beneficiaries. The result is an index that measures the prevalence of vote trading, and a network of vote traders. Its modular design provides great flexibility so that researchers can adapt it to other contexts by modifying these components.

\section*{Methodology}

We analyze a dataset covering more than 9 million votes over a period of 40 years of roll calls and bill (co)sponsorships in the U.S. Congress. Voting data comes from official roll call records published on the websites of the U.S. Senate and House (see Supplementary Material for a detailed description of the data as well as the data sources). For each chamber, we encode voting information into a matrix $\mathbb{V}$. Let entry $\mathbb{V}_{ik}=1$ if legislator $i$ voted Yes in roll call $k$ and $\mathbb{V}_{ik}=0$ if he or she voted No. $\mathbb{V}$ has dimension $N \times K$, where $N$ is the number of legislators and $K$ the number of roll calls. 

In line with \cite{stratmann_effects_1992,cohen_friends_2014} we assume that legislators' perceptions about the narrowness of vote outcomes can be approximated by the realized outcomes. Thus the narrower an observed vote outcome, the more likely is the presence of incentives to trade votes. However, narrow vote outcomes are rare and legislators arranging a trade of votes might prefer to arrange coalitions greater than the minimal winning coalition due to uncertainty as to other legislators' support \cite{weingast_rational_1979,groseclose_buying_1996}. Choosing the margin is thus a trade-off between the size of the set of potentially traded votes and the inclusion of only those votes posing strong incentives to trade. We define a narrow margin to be five or fewer votes and consider wide margin outcomes those that were passed or turned down with at least $20\%$ of the votes (see Supplementary Material for further details and robustness). In such clear outcomes we assume that legislators are likely to vote truthfully according to their preferred positions. We thus infer their preferred positions, or ideal points, as follows.

Consider a one-dimensional policy space where each bill can be located. Each legislator has an ideal point in this space, and so does each Yes or No vote on a particular bill (higher dimensional spaces can also be used if necessary, however, it has been shown that the first dimension captures over 70\% of the variation in voting decisions in the U.S. Congress \cite{poole_spatial_1985,poole_ideology_2007}). An important assumption in this kind of spatial voting models is that legislators suffer disutility the further their vote decision is from their ideal points in this policy space. Since wide margin roll calls are more likely to elicit sincere voting, we use them to estimate the ideal points in order to infer their preferred policy positions. For this purpose, we employ the Bayesian approach suggested by \cite{clinton_statistical_2004} and estimate the ideal point $\hat{{x}_i}$ of every legislator $i$ (see Supplementary Material for details on this step). Alternative methods \cite{poole_spatial_1985, porter_network_2005} can also be used but it has been shown that \cite{clinton_statistical_2004} differentiates better between extreme points. The ideal points allow us to compute the probability that legislator $i$ will vote Yes in roll call $k$. For this, we estimate for each roll call the probit model $\Pr(\mathbb{V}_{ik}=1|\hat{{x}_i})=\Phi(\hat{{x}_i}\beta_k)$, where $\Phi$ is the standard normal CDF and $\beta_k$ is the parameter to be estimated. We use this model to predict legislators' probabilities to vote Yes (see Supplementary Material for robustness to prediction errors). We collect these predictions in a matrix $\mathbb{Q}$ with dimensions $N \times K$. $\mathbb{Q}$ is key to detect deviations from preferred policy positions.

Let $\tau$ be a probability threshold to consider if a Yes vote is a deviation. For example, suppose that legislator $i$ votes Yes in a narrow roll call. However, he or she was predicted to vote No with $0.9$ probability, i.e. $\mathbb{Q}_{ik} = 0.1$. Then, $i$'s vote is considered a deviation only if $\mathbb{Q}_{ik} \leq \tau$. Of course, No votes can also be considered deviations, for example, if a legislator attempts to block the passage of a bill he or she is predicted to be in favor of. We focus on Yes votes because their beneficiaries can be deducted from different data (e.g. in (co)sponsorship data, constituency characteristics, or campaign finance data) and the large majority of theoretical work concentrates on them.

In order to identify the beneficiary of a deviation, we use (co)sponsorship data (which are assumed to approximate the legislator's preferences towards bills) based on the official bill data published by the Library of Congress. We encode this information into a matrix $\mathbb{S}$ with dimensions $N \times K$, where entry $\mathbb{S}_{ik}=1$ if legislator $i$ (co)sponsored the bill voted on in roll call $k$ and $\mathbb{S}_{ik}=0$ otherwise. Therefore, $i$'s vote in a narrow roll call $k$ is considered a directed deviation only if $\mathbb{V}_{ik}=1$, $\mathbb{Q}_{ik} \leq \tau$, and $\mathbb{S}_{jk}=1$ for some $j \neq i$. For our analysis, we assume that (co)sponsorships capture the explicit interests of legislators towards the passage of bills. However, other factors such as social ties or strong interests among the representatives' constituencies toward the bills could be encoded in $\mathbb{S}$ as well.

Now we can construct the \emph{directed deviation network} (DDN). In the DDN, an edge $i \rightarrow j$ indicates that legislator $i$ deviates in a roll call associated to a bill (co)sponsored by legislator $j$. The DDN is represented by the adjacency matrix $\mathbb{W}$, where entry $\mathbb{W}_{ij}>0$ if $i$ deviates one or more times to the benefit of $j$, and $\mathbb{W}_{ij}=0$ otherwise. We measure the reciprocity in $\mathbb{W}$ by adapting the method proposed by \cite{squartini_reciprocity_2013} to the context of logrolling. When defining reciprocity between legislators $i$ and $j$ as the symmetric part of their connections, expressed by $w_{ij}^{\leftrightarrow} = \min \left[\mathbb{W}_{ij} , \mathbb{W}_{ji} \right] = w_{ji}^{\leftrightarrow}$, the function

\begin{equation*}
    f_i(\mathbb{W}) = \begin{cases} 
      1 & \text{if } \sum_{j \neq i}^N w_{ij}^{\leftrightarrow} > 0 \\
      0 & \text{otherwise}
   \end{cases}
\end{equation*}
indicates whether legislator $i$ engaged in at least one reciprocal deviation. Then, the total number of representatives that reciprocate deviations is $R = \sum_i^N f_i(\mathbb{W})$ and the total number of deviations is $W = \sum_{i}^N \sum_{j \neq i}^N \mathbb{W}_{ij}$. The level of vote trading in the DDN is thus measured by $t = \frac{2R}{W}$, where $2R$ is a suitable predictor of the true number of votes traded. The total number of reciprocal edges is usually employed to measure network reciprocity \cite{katz_measurement_1955, achuthan_number_1982,wang_dyadic_2013,akoglu_quantifying_2012,squartini_reciprocity_2013,garlaschelli_patterns_2004}. In our context, however, we can show (based on a Monte Carlo Study) that the usually applied predictor inflates $t$ because it is significantly larger than the true number of trades. $2R$ is generally lower, so it corrects this bias (see SI for more details on the Monte Carlo simulation where we can control for each true trade).

The magnitude of $t$ might be driven by the topology of the DDN and thus by the data at hand. Therefore, we test the alternative hypothesis that deviations are the result of random errors rather than intentional behavior to benefit other legislators. For this test, we generate a sample of DDNs under the null model of random deviations. That is, we model each deviation as an independent Bernoulli random variable with probability of success $\mathbb{Q}_{ik}$. By simulating all the observed deviations in $\mathbb{V}$, we generate a null DDN. Repeating this procedure multiple times yields a sample of DDNs. We calculate the level of vote trading $t_0$ of each null DDN and compute the sample mean $\bar{t}_0$, which reflects the expected trading level under the null hypothesis. Following \cite{garlaschelli_patterns_2004,squartini_reciprocity_2013}, we construct the \emph{logrolling index}

\begin{equation}
    \ell = \frac{t - \bar{t}_0}{1 - \bar{t}_0}.\label{eq:index}
\end{equation}

The sign of $\ell$ is directly informative of the reciprocity in the DDN. If $\ell>0$, it means that legislators tend to deviate in favor of other legislators in a reciprocal fashion, suggesting vote trading. If $\ell<0$, then legislators tend to avoid reciprocity, indicating that the directed deviations are caused by other mechanisms. 

The final step is constructing the \emph{vote-trading network} (VTN). For this, we extract the reciprocal part of the DDN and remove pairs of reciprocal directed deviations that are not mutually profitable. Recall that the spatial model implies decreasing utility, and hence decreasing probability, of a Yes vote as the distance between the Yes position and the ideal points grows. Let $k_i$ and $k_j$ denote roll calls associated to bills that are (co)sponsored by $i$ and $j$ respectively. In addition, assume that $i$ deviated in $k_j$ and $j$ deviated in $k_i$. Profitability means that $i$'s Yes vote in $k_j$ yields more utility than his or her Yes vote in $k_i$. Therefore, trades are mutually profitable only if $\mathbb{Q}_{ik_i}<\mathbb{Q}_{ik_j}$ and $\mathbb{Q}_{jk_j}<\mathbb{Q}_{jk_i}$.

\section*{Results}

\begin{figure}[htb!]
\centering
\caption{Empirical relationship between $\tau$ and $\ell$. The index was computed by generating samples of 10,000 null DDNs for each level of $\tau$. Standard errors of $\ell$ were computed using the Jackknife procedure proposed by \cite{newman_mixing_2003}.}
\includegraphics[width=1\textwidth]{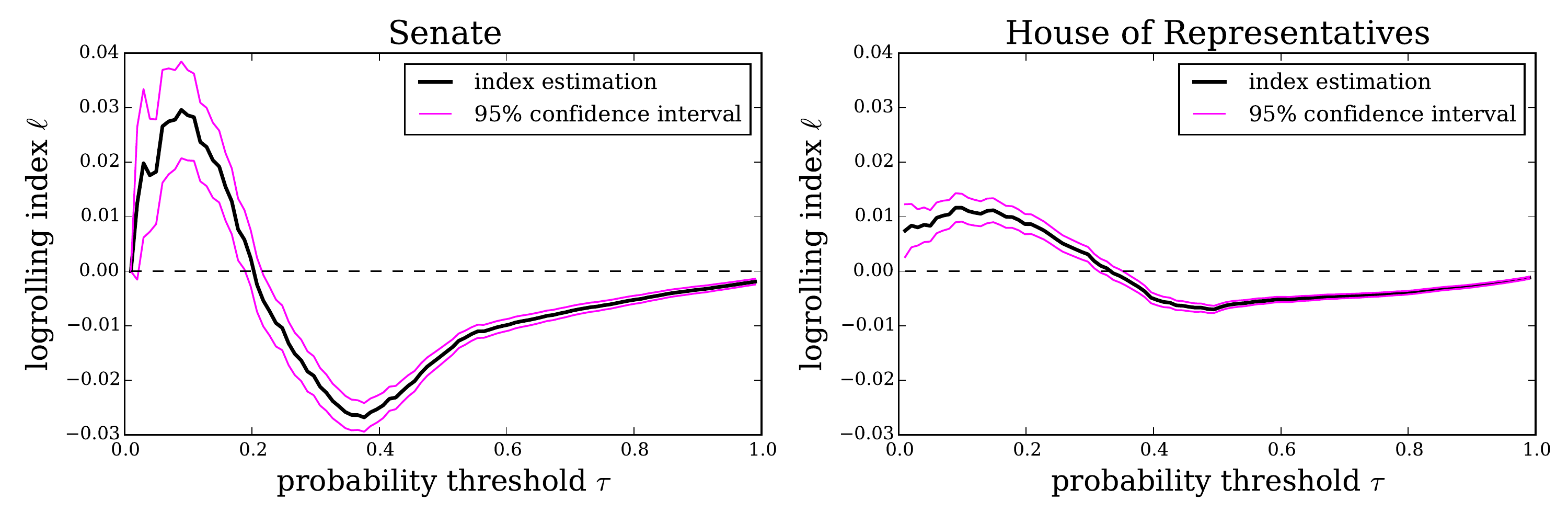}
\label{fig:tauVsEll}
\end{figure}

\begin{figure}[htb]
\centering
\caption{Vote-trading network of the Senate. Nodes are colored according to the party of longest affiliation. Their sizes are proportional to their degree. If the color of an edge does not match the color of its nodes (e.g. blue edge between red nodes), it is because at least one of the nodes had a party affiliation at the moment of the vote that was different from the one captured by its color. We recommend this graphic is viewed digitally (rather than in hard copy format) to allow easy magnification of legislators' names.}
\includegraphics[angle=0,width=1\textwidth]{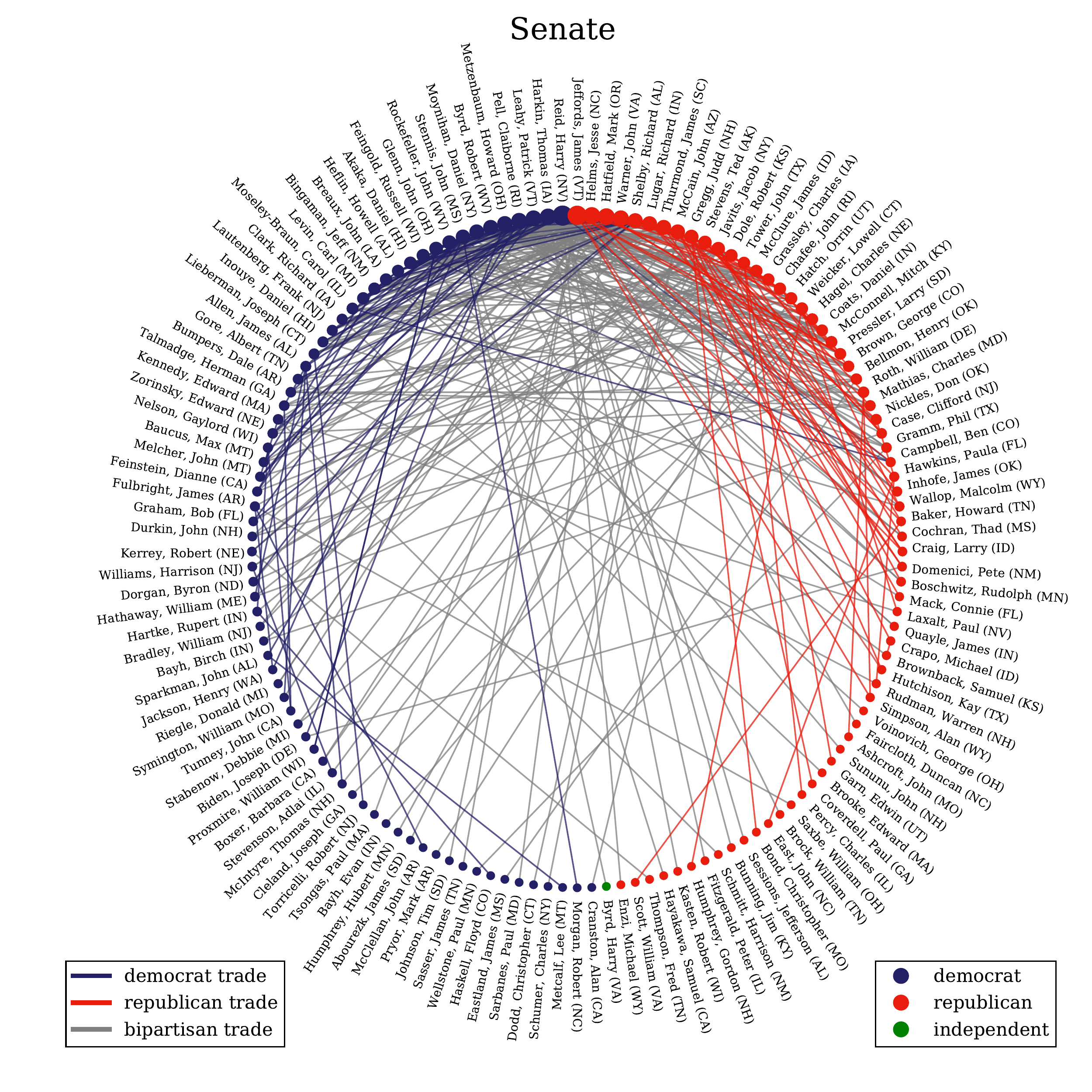}
\label{fig:VTNs}
\end{figure}

Figure~\ref{fig:tauVsEll} shows the result of estimating \eqref{eq:index} for different values of $\tau$ in the two chambers. In both cases, $\ell$ is positive and statistically significant for lower values of $\tau$. This result is intuitive because it suggests that vote trading can only be detected when the probability of a Yes is lower than that of a No. In other words, if we would expect legislators to vote Yes most of the time, then the DDN would be full of false deviations. Since we are interested in gaining as much information as possible about the VTN of each chamber, we choose for the following analyses the largest possible value of $\tau$ such that $\ell$ is positive and statistically significant. Figure~\ref{fig:VTNs} shows the VTNs of the Senate and the House. The graphs reveal that, while trades among members of the same party are common, a large part of vote trading is bipartisan. This is true for both the Senate and the House.

Table~\ref{tab:summary} shows the summary statistics of the VTNs from the Senate and the House with respect to party affiliation and network characteristics. In addition, the table lists the congressmen with most trades in the VTNs as well as the top coalitions (see Supplementary Material for the complete lists). These results are in line with qualitative studies and theoretical work on vote trading. For example, it has been argued that Senators are relatively more engaged in vote trading than Representatives and, more generally, that longer tenureships in office might be a central factor for stable vote-trading relationships \cite{matthews_u.s._1973,bernholz_stability_1978}. The average level of the logrolling index in its positive region is significantly higher in the Senate.  Relatively more Senators participated in trades than Representatives, and relatively more roll calls (and more bills) were affected in the Senate than in the House. In addition, the average number of trades per dyad is higher in the Senate, suggesting more intensity in vote-trading activity. Lengthier time between trades in the Senate is consistent with the idea that longer relationships between Senators (fostered by the institution of 6-year terms) are important in building up trust in order to effectively engage in vote trading.

\begin{figure}[htb]
\centering
\caption{Vote trading, party division, and polarization in the 93rd to the 113th U.S. Congress. The top panel shows the number of trades between members of the same party as well as bipartisan trades. The middle panel shows the shares of Republican and Democratic seats, with a horizontal line indicating 50\% of the seats. The bottom panel depicts political polarization in the respective chamber (based on Voteview.com's broadly applied polarization measure \cite{poole_spatial_1985}, normalized to the value of the 113th Congress). The dotted orange vertical line indicates the congress during which the Tea Party movement evolved.}
\includegraphics[angle=0,width=1\textwidth]{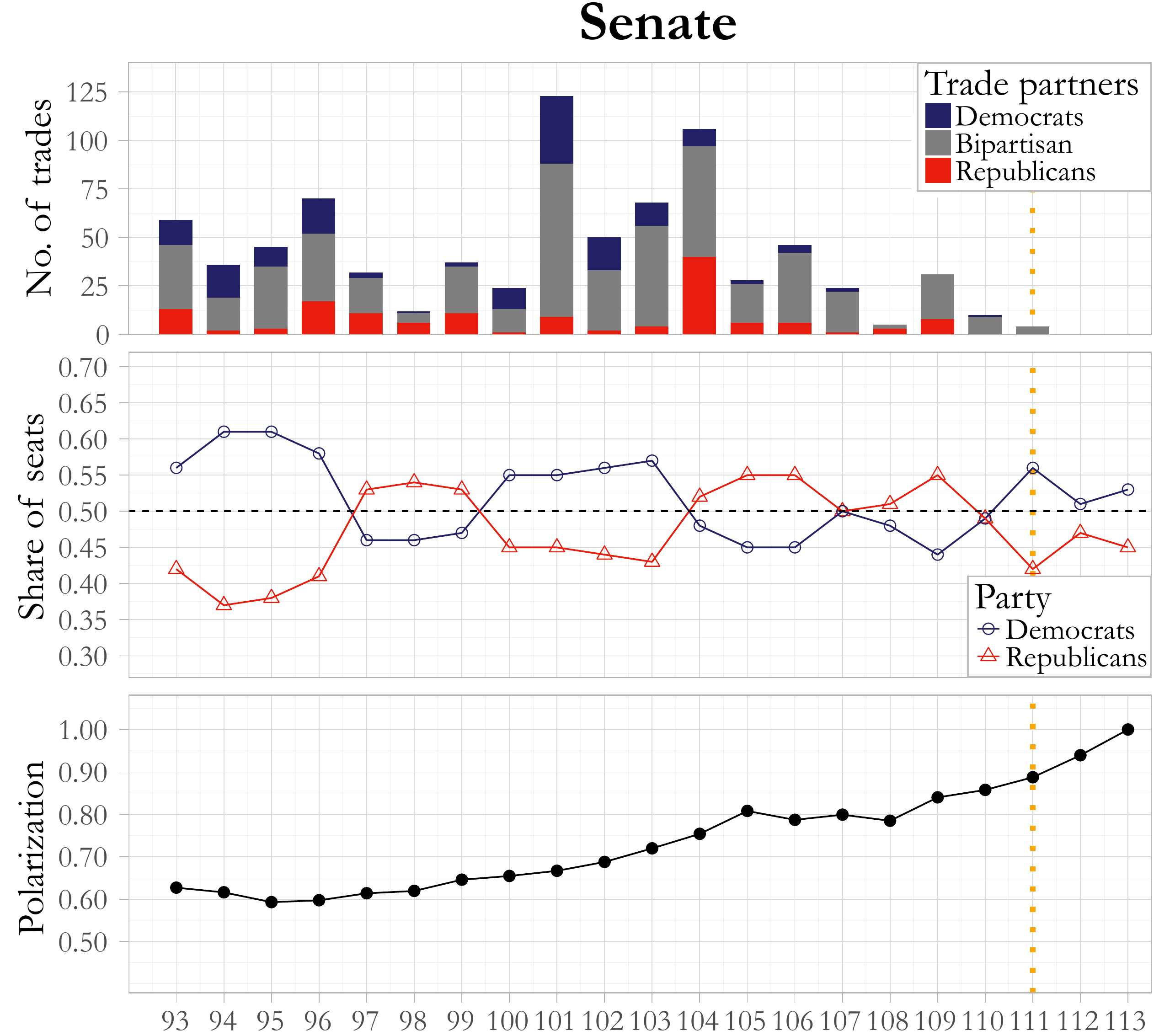}
\label{fig:partydiv}
\end{figure}

Figure~\ref{fig:partydiv} depicts how logrolling in the Senate and House has evolved over the last four decades with respect to two potentially important drivers of vote trading: party division and political polarization \cite{poole_spatial_1985}. The figure suggests that vote trading is common in most congresses but varies substantially over time. Within-party trading is particularly present in, but not restricted to, the majority party. Interestingly, bipartisan cooperation in the form of vote trading is a common feature of the political process in both chambers during most congresses. Moreover, the share of bipartisan trades is substantially higher in the Senate compared to the House. This finding is in line with the widespread view that party politics is less relevant in the Senate compared to the House \cite{monroe_why_2008}. In recent congresses (with very high levels of polarization) both partisan and bipartisan trading has strongly declined. Note that the decrease in Republican trades stands in stark contrast to previous congresses with Republican majorities, indicating that inner-party struggles related to the advent of the Tea Party Caucus within the Republican Party might have substantially hindered within-party cooperation \cite{kabaservice_rule_2013}.

\begin{figure}[htb!]
\centering
\caption{The VTNs were treated as undirected unweighted graphs because all trades are reciprocal and weights are negligible (see Table~\ref{tab:summary}). We fitted the data by fixing the theoretical shape parameter $\alpha=0.5$ and estimated $\beta=10.6$ for the Senate and $\beta=14.8$ for the House via MLE. We generated synthetic data from these distributions and performed two-sample Anderson-Darling tests. The theoretical model cannot be rejected in any chamber.}
\includegraphics[width=1.\textwidth]{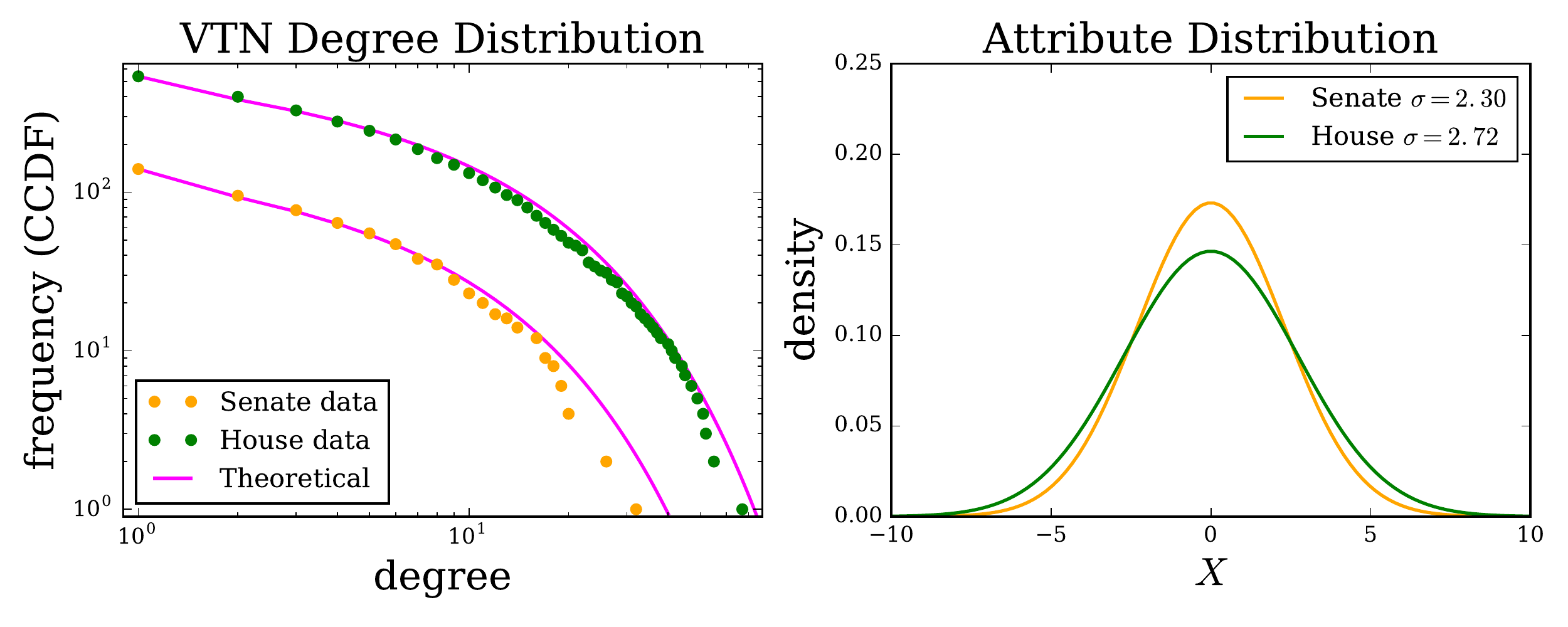}
\label{fig:fitAndModel}
\end{figure}

The topology of the VTN is informative about the concentration of trades among legislators. The left panel in Figure~\ref{fig:fitAndModel} presents both degree distributions. Interestingly, both chambers exhibit a Gamma distribution

\begin{equation}
    \Pr(x; \alpha, \beta) = \frac{1}{\Gamma(\alpha)\beta^\alpha} x^{\alpha-1} e^{-\frac{x}{\beta}}\label{eq:gamma}.
\end{equation}
It is remarkable that the degrees of these independent VTNs are well explained by this distribution. Furthermore, the null hypothesis fails to generate VTNs with Gamma-distributed degrees (see Supplementary Material), confirming that these findings are a feature of the data rather than an artifact of our methodology.

These results can be explained by an underlying random variable $X$, distributed across legislators according to a normal $N(0,\sigma)$ (right panel in Figure~\ref{fig:fitAndModel}). $X$ may represent a factor that causes some legislators to be more successful than others at accomplishing successful vote trades. We represent this mechanism by $Y = X^2$, where $Y$ is the degree. Then, by the properties of sums of squared normal random variables, $Y \sim \text{Gamma}\left(\frac{1}{2}, 2\sigma^2\right)$. Aderson-Darling tests fail to reject that the empirical data comes from this distribution (see Figure~\ref{fig:fitAndModel}).

This model can be used to test candidate explanations of the distribution of trades across legislators, for example, social skills \cite{deming_growing_2015}. It can also be extended to investigate logrolling from a network perspective, taking into account contextual knowledge about the various motivations why specific legislators would choose to trade with each other. The specifics about who connects with whom can be modeled through, for example, different variations of the configuration model \cite{molloy_critical_1995}, or agent-computing models \cite{axelrod_complexity_1997,axtell_why_2000}.

\section*{Conclusions}

The method presented combines well-established theory from Political Science, Economics, and Game Theory with big data and network science (both of which are considered increasingly relevant for new scientific discoveries in Political Science and Economics \cite{fowler_connecting_2006, lazer_computational_2009, einav_economics_2014, jackson_networks_2014, cranmer_kantian_2015}). However, it has some limitations that could be addressed in future research. Although we can test the statistical significance of logrolling at the aggregate level, we cannot tell whether all the reciprocal deviations revealed in the VTN are true trades (hence we use $2R$ to reduce false positives). Any conclusions with respect to low-degree legislators in the VTN thus need to be considered with care. On a more general level, it is important to keep in mind that there might well be other forms of vote trading. Three aspects are of particular relevance in this context. First, (co)sponsorships are not the only signal of preferences towards bills. Second, it might well be that some deals are arranged between groups. In those cases, our method would likely capture the roll call/bill affected by such trades, but not each individual participating. Third, our method does not capture `implicit' vote trading (i.e., trades in the form of issue-packages or `platforms' \cite{tullock_simple_1970} and exchanging favors in order to support or oppose policies before they can officially be voted on). 

Previous research has taken an econometric approach (like the seminal work by \cite{stratmann_effects_1992}) that is hardly scalable and relies on ex-ante knowing the specific socioeconomic context of bills and legislators where vote trading is suspected. In contrast, our method is designed to study general aspects of logrolling on a large scale. The approach is flexible, scalable, and provides results in a very intuitive and informative manner (e.g. through VTNs), making it easily applicable to various legislative assemblies and time frames. In fact, with more voting and sponsorship data available, our method can be applied to study hidden cooperation in the form of vote trading across U.S. states, different counties, as well as different types of assemblies. While far from conclusive, our results present first insights into vote trading as a means to potentially overcome political obstruction generated by increasing ideological polarization. However, such hidden cooperation in the presence of increasing polarization is only possible up to a certain point of ideological cleavage. Insights in this direction are important in informing the design of democratic institutions.  More broadly, the ability to detect and measure hidden cooperation such as logrolling might reveal important insights about the dynamics underpinning hidden human cooperation in general.


\bibliographystyle{IEEEtran}

\clearpage

\begin{figure}[htb]
\centering
\includegraphics[angle=0,width=1\textwidth]{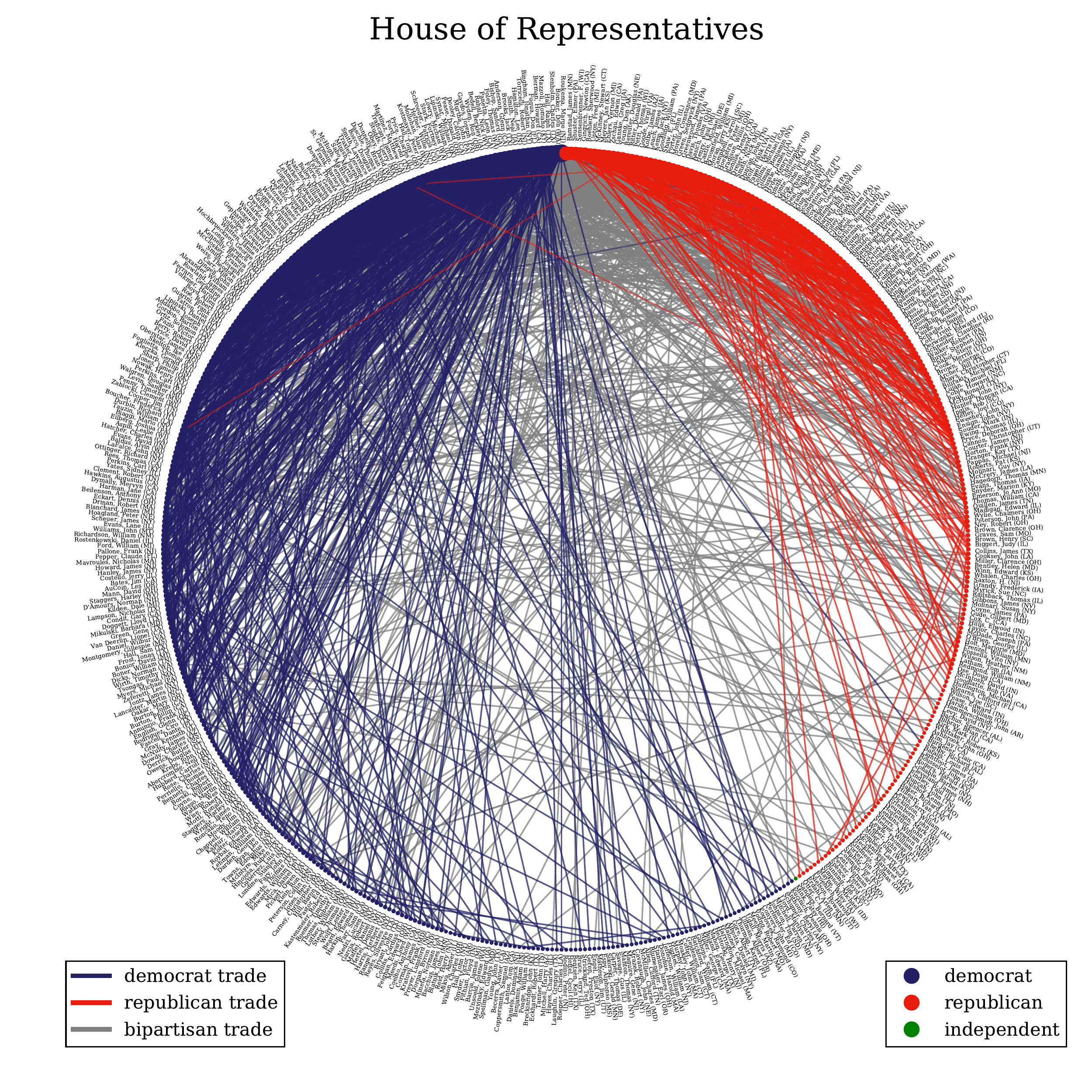}
\end{figure}

\begin{figure}[htb]
\centering
\includegraphics[angle=0,width=1\textwidth]{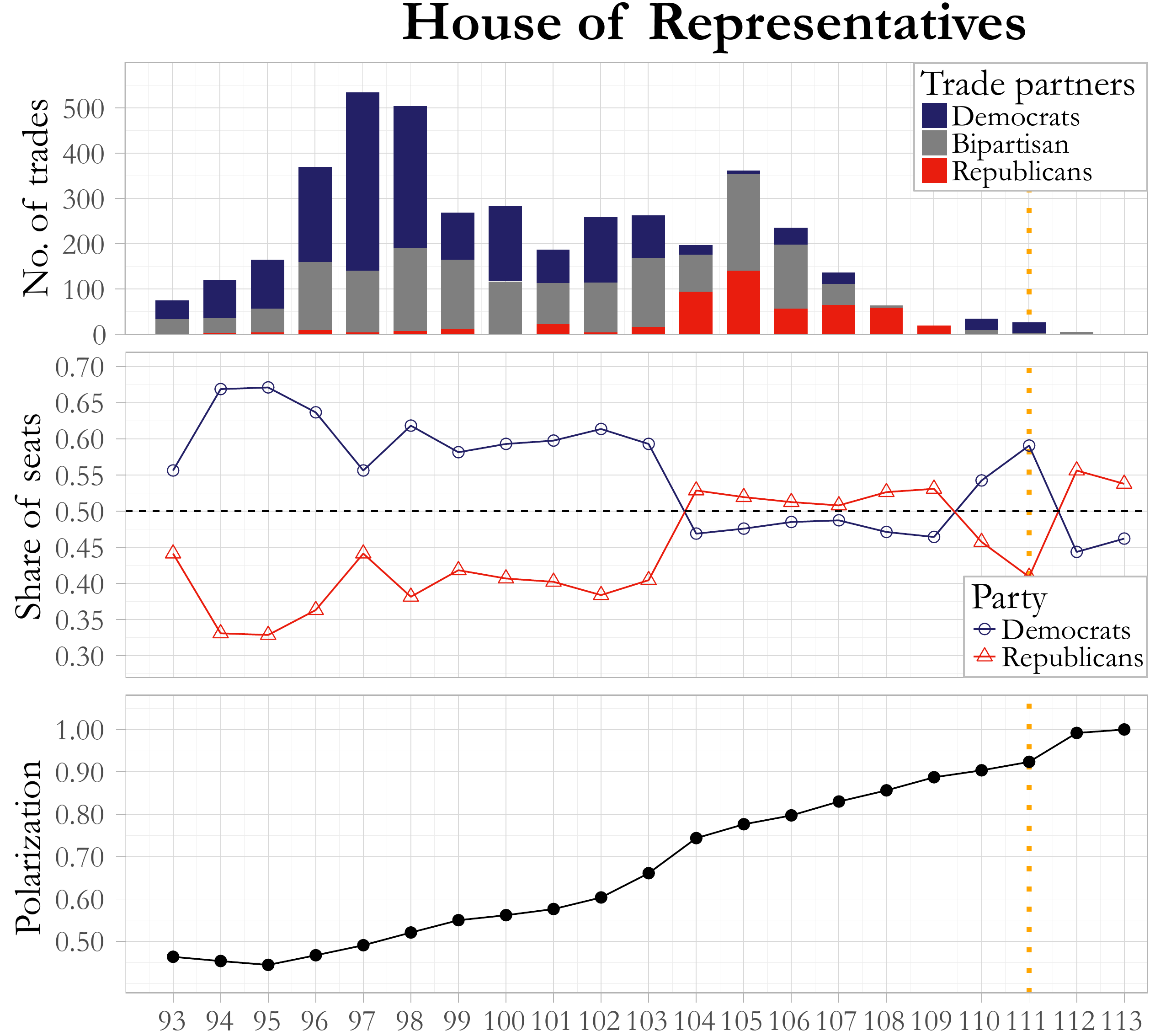}
\end{figure}

\clearpage

\begin{table}[htb!]
\centering
\caption{VTN summary statistics\label{tab:summary}}
\begin{tabular}{lrr}
\hline
\hline
Attribute & Senate & House\\
\hline
Average reciprocity & 0.019 $\pm$ 0.008 & 0.008 $\pm$ 0.002 \\
Legislators & 140 (37.14\%)  & 538 (31.17\%)  \\
By party: Dem. (Rep.) [Ind.] & 71 (68)  [1]  & 320 (217)  [1]  \\
Bills & 173 (5.97\%) & 228 (2.13\%) \\
Votes & 810 (0.07\%) & 4,114 (0.05\%) \\
\hline
Average degree & 11.57 & 15.29 \\
Average trading partners & 5.27 & 7.41 \\
Highest degree (partners) & 74 (32) & 136 (67) \\
Average trades per partnership & 2.20 & 2.07 \\
Average years between trades & 4.78 & 4.21 \\
\hline
Most central party (votes traded) & Democrat (416) & Democrat (2,737) \\
Democrat trades & 157 & 1,874 \\
Republican trades & 143 & 522 \\
Bipartisan trades & 510 & 1,718 \\
\hline
\end{tabular}
\vspace{1ex}\\
\raggedright{Average reciprocity is the mean of $\ell$ in its positive and statistically significant region. We report its average 95\% confidence interval. The average number of years between trades was calculated by averaging the difference between the dates of all possible exchanges of each dyad from the DDN. This was necessary because in reality we do not know which precise vote was traded for which other.}
\end{table}

\end{document}